\documentclass[preprint,aps]{revtex4}
\usepackage{graphicx}

\begin{document}

\title{Equivalence of the measures of
non-Markovianty for open two-level systems}
\author{Hao-Sheng Zeng\footnote{Corresponding author
email: hszeng@hunnu.edu.cn}, Ning Tang, Yan-Ping Zheng and Guo-You
Wang}\affiliation{Key Laboratory of Low-Dimensional Quantum
Structures and Quantum Control of Ministry of Education, and
Department of Physics, Hunan Normal University, Changsha 410081,
China}

\begin{abstract}
In order to depict the deviation of quantum time evolution in open
systems from Markovian processes, different measures have been
presented. We demonstrate that the measure proposed by Breuer, Laine
and Piilo [Phys. Rev. Lett. \textbf{103}, 210401 (2009)] and the two
measures proposed by Rivas, Huelga and Plenio [Phys. Rev. Lett.
\textbf{105}, 050403 (2010)] have exactly the same non-Markovian
time-evolution intervals and thus are really equivalent each other
when they apply to open two-level systems coupled to environments
via Jaynes-Cummings or dephasing models. This equivalence implies
that the three measures in different ways capture the intrinsical
characters of non-Markovianty of quantum evolutional processes. We
also show that the maximization in the definition of the first
measure can be actually removed for the considered models without
influencing the sensibility of the measure to detect
non-Markovianty.

\end{abstract}

\maketitle

\section{Introduction}
The evolution of open quantum systems can be divided into two basic
types, Markovian and non-Markovian processes \cite{Breuer3}. For
Markovian processes, the correlation time between the system and
environment is considered to be infinitesimally small so that the
dynamical map does not carry any memory effects, leading to a
monotonic flow of the information from system to environment. In
contrast, non-Markovian processes with memory have different
dynamical traits which give rise to the back flow of information
from environment to the system \cite{Breuer,Laine}. Recently, people
found that non-Markovian processes can lead to distinctly different
effects on decoherence and disentanglement
\cite{Dijkstra,Anastopoulos} of open systems compared to Markovian
processes. Many relevant physical systems, such as quantum optical
system \cite{Breuer3}, quantum dot \cite{Kubota}, color-core spin in
semiconductor \cite{Kane}, could not be described simply by
Markovian dynamics. In addition, quantum chemistry \cite{Shao} and
excitation transfer of biological system \cite{Chin} also need to be
treated as non-Markovian processes. Because of these distinct
properties and extensive applications, more and more attention and
interest have been devoted to the study of non-Markovian processes
of open systems, including the measures of non-Markovianty
\cite{Breuer,Laine,Rivas,Wolf,Usha,Lu}, the positivity
\cite{Breuer1,Shabani,Breuer2}, and some other dynamical properties
\cite{Haikka,Chang,Krovi,Chru} and approaches \cite{Jing,Koch} of
non-Markovian processes. Experimentally, the simulation
\cite{Xu1,Xu2} of non-Markovian environment has been realized.

The measure of non-Markovianty of quantum evolution is a fundamental
problem which aims to detect whether a quantum process is
non-Markovian and how much degrees it deviates from a Markovian one.
So far, almost all measures can only serve to settle the former,
i.e., serve as the sufficient (not necessary) condition for the
emergence of non-Markovianty. Therefore, the problem for measuring
the non-Markovianty of quantum processes still remains elusive and,
in some sense, controversial. Based on the distinguishability of
quantum states, Breuer, Laine and Piilo (BLP) \cite{Breuer} proposed
a measure to detect the non-Markovianty of quantum processes which
is linked to the flow of information between system and environment.
Alternatively, Rivas, Huelga and Plenio (RHP) \cite{Rivas} also
presented two measures of non-Markovianty by exploiting the
dynamical behavior of quantum entanglement under the local
trace-preserving CP maps. Other measures of non-Markovianty include
the one proposed by Wolf \emph{et al.} \cite{Wolf} based on the
breakdown of the semigroup property, the one proposed by Lu \emph{et
al.} \cite{Lu} using quantum Fisher information flow, and the ones
proposed by Usha Devi \emph{et al.} \cite{Usha} using relative
entropy difference and fidelity difference. It is worthwhile to
stress that these measures of non-Markovianity are generally not the
same. Thus, studying and exposing the relations among all these
measures under some specific models becomes very important. Haikka
\emph{et al.} \cite{Haikka1} studied the links between the BLP
measure and one of the RHP measures very recently, for a
laser-driven qubit system embedded in a structured Lorentzian
environment. By algebraic plus numerical calculations, they showed
that, in the nonsecular regime, the two measures agree. But for
other more cases, no definite result can be presented.

In this paper, we study the equivalence of the three measures of
non-Markovianity recently proposed by BLP \cite{Breuer} and by RHP
\cite{Rivas} in the case that a two-level system coupling to its
environment via damped Jaynes-Cummings or dephasing model. Two
important results are found: Firstly, the three measures are exactly
equivalent in the aspect of detecting non-Markovianty for the
involved models. And secondly, for our considered models, we find
that the maximization in the definition of the BLP measure may be
actually removed. This solves the problem that is being explored
extensively by many researchers \cite{Xu,He}. Our researches have
the following features. First, the interaction models involved here
are most fundamental in the theoretical studies of dynamics of open
quantum systems, and we need not assuming any specific spectral
density for the structured environment. Thus, our results possess
good adaptability. Next, our operations are completely analytical
and hence the results are more convincing.

The article is organized as follows. In Sec. II we briefly review
the three measures proposed by BLP and RHP respectively. The
definition of equivalence between different measures is also
presented. And in Secs. III and IV we study the equivalence of the
three measures for a two-level system interacting with environment
via damped Jaynes-Cummings model and dephasing model respectively. A
conclusion is arranged in Sec.V.

\section{Review of the three measures}
In reference \cite{Breuer}, BLP presented a measure of
non-Markovianty for quantum processes of open systems which is based
on the idea that Markovian processes tend to continuously reduce the
trace distance between any two states of a quantum system. Thus, an
increase of trace distance during any time intervals implies the
emergence of non-Markovianty. The authors further linked the changes
of trace distance to the flow of information between system and its
environment, and concluded that the back flow of information from
environment to the system is the key feature of a non-Markovian
dynamics. Considering that the measure should reveal the total
feature of a whole quantum process, they thus suggest the quantity,
\begin{equation}
\mathcal{N}=\max_{\rho_{1,2}(0)}\int_{\sigma>0}dt\sigma(t,\rho_{1,2}(0)),
\end{equation}
as the measure of non-Markovianty of a quantum process. Where
$\sigma(t,\rho_{1,2}(0))$ denotes the time derivative of trace
distance for a pair of dynamical states with initial values
$\rho_{1,2}(0)$ of the considered system. The time integration is
extended over all intervals in which $\sigma$ is positive, and the
maximum is taken over all pairs of initial states. For any Markovian
process, $\mathcal{N}=0$. And if $\mathcal{N}>0$, the process must
be non-Markovian.

Alternatively, RHP \cite{Rivas} presented another method to measure
the non-Markovianty of a quantum process which is based on the
monotonic drop of quantum entanglement between bipartite systems
under the influence of local Markovian environments. Suppose a
system of interest is initially prepared in a maximally entangled
state with an ancillary particle, where only the system is
influenced by a noise environment and the ancillary particle is
free. Then the quantity for measuring the non-Markovianty of the
quantum process is defined as
\begin{equation}
\mathcal{I}^{(E)}=\int_{t_{0}}^{t_{max}}|\frac{dE[\rho_{SA}(t)]}{dt}|dt-\triangle
E,
\end{equation}
where the time derivative in the integrand is for the dynamical
entanglement between the system and ancillary particle, $\triangle
E=E[\rho_{SA}(t_{0})]-E[\rho_{SA}(t_{max})]$ denotes the difference
of entanglements at the initial time $t_{0}$ and the final time
$t_{max}$ of the interest quantum process.

The above two measures, $\mathcal{N}$ and $\mathcal{I}^{(E)}$, are
introduced respectively through the monotonicity of trace distance
or quantum entanglement under CP (or local CP) maps. Their
mathematical forms are similar actually. It is not difficult to find
that eq.(2) may be equivalently rewritten as
\begin{equation}
\mathcal{I}^{(E)}=2\int_{\dot{E}(t)>0}\dot{E}(t)dt,
\end{equation}
with $\dot{E}(t)$ being the time derivative of dynamical
entanglement $E(t)$. Obviously, the measure $\mathcal{I}^{(E)}$ is
very similar to $\mathcal{N}$ in forms, with the time derivative
$\dot{E}(t)$ of dynamical entanglement replaced by the time
derivative $\sigma=\dot{D}$ of the dynamical trace distance
$D(\rho_{1}(t),\rho_{2}(t))$. The visible difference is: the measure
$\mathcal{N}$ involves a maximum over all pairs of initial states,
while $\mathcal{I}^{(E)}$ is defined via a given maximally entangled
initial state and thus escapes from the optimization problem.
Actually, the escaping of the optimization is just one of the
original intentions for the measure $\mathcal{I}^{(E)}$ to be
proposed. In the following sections, we will demonstrate that, for
the models considered in this article, the maximization process can
be removed actually.

The third quantity for measuring the non-Markovianty of a quantum
process is also based on the entanglement dynamics between the
system and an ancillary particle. Given the maximally entangled
state $|\Phi\rangle$ of the system plus the ancillary particle, a
locally complete positive map $\varepsilon$ will keep the positivity
of density operator $\rho=|\Phi\rangle\langle\Phi|$ invariable,
i.e., $\varepsilon (|\Phi\rangle\langle\Phi|)\geq0$. Starting from
this point of view, a measure of non-Markovianty of quantum
processes is then born \cite{Rivas}

\begin{equation}
\mathcal{I}=\int_{0}^{\infty}g(t)dt,
\end{equation}
with
\begin{equation}
g(t)=\lim_{\epsilon\rightarrow0^{+}}\frac{\|[I+(\mathcal{L}_{t}\otimes
I)\epsilon]|\Phi\rangle\langle\Phi|\|-1}{\epsilon}.
\end{equation}
Where $\mathcal{L}_{t}$ is the super operator in the non-Markovian
dynamical master equation $\frac{d\rho}{dt}=\mathcal{L}_{t}(\rho)$
for the open system.

The above three measures $\mathcal{N}$, $\mathcal{I}^{(E)}$ and
$\mathcal{I}$, for the non-Markovianty of a quantum process are
introduced through different ways. The physical meanings are
different. A question naturally arises: They are equivalent each
other? In this paper, we demonstrate that the three measures are
equivalent each other when they apply to open two-level systems with
damped Jaynes-Cummings or dephasing models. This equivalence implies
that the measures, from different sides, well capture the
intrinsical characters of non-Markovianty of quantum evolutional
processes.

As the end of this section, let us elaborate the implication for
different measures to be equivalent. We say two measures $M_{1}$ and
$M_{2}$ to be equivalent means that: For a given quantum process, if
by measure $M_{1}$, the non-Markovianty emerges, then it also
emerges by measure $M_{2}$; Conversely, if by measure $M_{1}$, the
non-Markovianty does not emerge, it also does not by measure
$M_{2}$. The exact logics are: $M_{1}>0 \Longleftrightarrow
M_{2}>0$, $M_{1}=0 \Longleftrightarrow M_{2}=0$.

\section{The case of Jaynes-Cummings model}
In this section, we will demonstrate the equivalence of the three
measures of non-Markovianty for the case of a two-level system
coupled to its environment via Jaynes-Cummings model. For this end,
we should first derive out the expressions of the three measures
given the specific interaction model. For the calculation of
$\mathcal{N}$, let us consider a two-level atom interacting with its
environment via a Jaynes-Cummings model. The environment is assumed
to be initially in the vacuum state, but its spectral density is
arbitrary. This case can be solved exactly. For any pair of initial
atomic states, $\rho_{1}(0)$ and $\rho_{2}(0)$, one can obtain the
time derivative of trace distance of the corresponding dynamical
states as\cite{Laine},

\begin{equation}
\sigma(t,\rho_{1,2}(0))=\frac{2|G(t)|^{2}a^{2}+|b|^{2}}{\sqrt{|G(t)|^{2}a^{2}+|b|^{2}}}\frac{d}{dt}|G(t)|,
\end{equation}
where
$a=\langle1|\rho_{1}(0)|1\rangle-\langle1|\rho_{2}(0)|1\rangle$ and
$b=\langle1|\rho_{1}(0)|0\rangle-\langle1|\rho_{2}(0)|0\rangle$ are
the differences of populations and of coherences respectively
between the two given initial states. The function $G(t)$ is defined
as the solution of the integro-differential equation,
\begin{equation}
\frac{d}{dt}G(t)=-\int_{0}^{t}dt_{1}f(t-t_{1})G(t_{1}),
\end{equation}
with initial condition $G(0)=1$. The kernel $f(t-t_{1})$ denotes the
two-point reservoir correlation function which is the Fourier
transformation of the spectral density. Introducing the
time-dependent decay rate,

\begin{equation}
\gamma(t)=-2\mathrm{Re}[\frac{\dot{G}(t)}{G(t)}]=-\frac{2}{|G(t)|}\frac{d}{dt}|G(t)|,
\end{equation}
then the derivative of the trace distance may be rewritten as,

\begin{equation}
\sigma(t,\rho_{1,2}(0))=-\gamma(t)F(t).
\end{equation}
Here the completely positive real function $F(t)$ is defined as,

\begin{equation}
F(t)=\frac{a^{2}e^{-\frac{3}{2}\Gamma(t)}+|b|^{2}e^{-\frac{1}{2}\Gamma(t)}}
{\sqrt{a^{2}e^{-\Gamma(t)}+|b|^{2}}},
\end{equation}
with $\Gamma(t)=\int_{0}^{t}dt'\gamma(t')$. According to eq.(1), the
non-Markovian measure $\mathcal{N}$ hence may be expressed as,
\begin{equation}
\mathcal{N}=-\int_{\gamma(t)<0}\gamma(t)F(t)dt.
\end{equation}
Where we assume that the pair of initial states $\rho_{1,2}(0)$ just
maximizes $\mathcal{N}$ so that the maximization symbol is removed.
Actually, due to the completely positivity of function $F(t)$, the
time intervals in which the trace distance increases monotonously,
or equivalently the intervals in which non-Markovianty emerges (we
call them non-Markovian intervals below), are uniquely determined by
the condition $\gamma(t)<0$. The change of initial states
$\rho_{1,2}(0)$ would not alter the positions and lengthes of these
non-Markovianty intervals, i.e., would not alter the distribution of
them. No matter how the initial states chnage, $\mathcal{N}$ would
not shift from positive to zero, or vice versa. In this sense, the
maximization to eq.(11) may be removed without influencing the
sensibility of $\mathcal{N}$ to detect non-Markovianty. This is an
important result which solves the problem being explored by many
researchers \cite{Xu,He}.

Next, we begin with the calculation of the second measure
$\mathcal{I}^{(E)}$. Consider a maximally entangled state of two
two-level atoms,

\begin{equation}
|\Phi\rangle=\frac{1}{\sqrt{2}}(|10\rangle_{SA}+|01\rangle_{SA}),
\end{equation}
where atom 1 is our considered system which couples to an
environment via Jaynes-Cummings interaction and atom 2 is an
ancillary particle which remains isolated from environment. The
whole Hamiltonian reads,

\begin{eqnarray}
H &=& H_{0}+H_{I}, \\
\nonumber
H_{0} &=& \sum_{i=1}^{2}\omega_{0}\sigma_{+}^{(i)}\sigma_{-}^{(i)}+ \sum_{k}{\omega_{k}{a_{k}}^{\dagger}a_{k}},\\
\nonumber H_{I} &=& \sum_{k}g_k\sigma_{+}^{(1)}a_k+h.c.,
\end{eqnarray}
where $\sigma_{+}^{(i)}$ and $\sigma_{-}^{(i)}$ denote ladder
operators for the $i$-th atom, $\omega_{k}$ and $a_{k}$ are the
frequency and annihilation operator for the $k$-th harmonic
oscillator of the environment, $g_{k}$ is the coupling constant. We
assume the two atoms have the same transition frequency
$\omega_{0}$. Introducing to the interaction picture with respect
$H_{0}$, one has,

\begin{equation}
\tilde{H}_I=\sigma_{+}^{1}\sum_{k}{{g_k}{a_k}\mathrm{exp}[{(\omega_0-\omega_k)t}]}+h.c.
\end{equation}
Assume that the two atoms are initially in the entangled state
$|\Phi\rangle$ and environment is initially in vacuum state
$|0\rangle$, then the dynamical wave function for the compound
system including both atoms and environment may be expressed as,

\begin{equation}
|\psi(t)\rangle={c_1(t)}|10\rangle_{SA}|0\rangle_E+{c_2(t)}|01\rangle_{SA}|0\rangle_E+\sum_k{{c_k(t)}|00\rangle_{SA}|1_k\rangle_E}.
\end{equation}
The dynamical reduced state for the two atoms thus reads,

\begin{equation}
\begin{array}{c}
{\rho}_{SA}(t)=(1-|c_{1}(t)|^{2}-|c_{2}(t)|^{2})|00\rangle\langle00|+|c_{1}(t)|^{2}|10\rangle\langle10|
+|c_{2}(t)|^{2}|01\rangle\langle01| \\
+c_{1}(t)c_{2}^{*}(t)|10\rangle\langle01|+c_{1}^{*}(t)c_{2}(t)|01\rangle\langle10|,
\end{array}
\end{equation}
where the time-dependant coefficients are governed by
Schr\"{o}dinger equation,

\begin{equation}
-i\frac{\partial}{\partial{t}}|\psi(t)\rangle=\tilde{H}_I(t)|\psi(t)\rangle.
\end{equation}
After a straightforward deduction, we find
$c_{2}(t)=c_{2}(0)=1/\sqrt{2}$ and $c_{1}(t)=G(t)/\sqrt{2}$ with
$G(t)$ being determined by eq.(7). If we use concurrence to describe
the entanglement between the system and ancillary particle, we have
from eq.(16) that,

\begin{equation}
C({\rho}_{SA}(t))=2|c_1^*(t)c_2(t)|=|G(t)|,
\end{equation}
and thus,
\begin{equation}
\frac{d}{dt}C(\rho_{SA}(t))=-\frac{1}{2}\gamma(t)e^{-\Gamma(t)/2},
\end{equation}
here the definition of $\gamma(t)$ in eq.(8) is used. Eq.(5) then
immediately produces,
\begin{equation}
\mathcal{I}^{(E)}=-\int_{\gamma(t)<0}\gamma(t)e^{-\Gamma(t)/2}dt.
\end{equation}
It again shows that the non-Markovian intervals for measure
$\mathcal{I}^{(E)}$ are uniquely determined by condition
$\gamma(t)<0$.

For the calculation of the third measure $\mathcal{I}$, we need the
dynamical master equation of the open two-level system
\cite{Breuer3},

\begin{equation}
\frac{d{\rho}}{dt}=-\frac{i}{2}S(t)[\sigma_{+}\sigma_{-},\rho]+
\gamma(t)[{\sigma}_-{\rho}{\sigma}_+-\frac{1}{2}\{{{\sigma}_+{\sigma}_-,{\rho}}\}],
\end{equation}
here $S(t)=-2\mathrm{Im}(\frac{\dot{G}(t)}{G(t)})$ with $G(t)$
determined by eq.(7) and $\gamma(t)$ is defined by eq.(8). After a
straightforward calculation, one find from eq.(4) that
\begin{equation}
g(t)=\left\{%
\begin{array}{lll}
    0 & \mathrm{for} &\gamma(t)\geq0 \\
    -\gamma(t) &\mathrm{for} & \gamma(t)<0\\
\end{array}%
\right.
\end{equation}
and thus
\begin{equation}
\mathcal{I}=-\int_{\gamma(t)<0}\gamma(t)dt.
\end{equation}
Once again, the non-Markovian intervals for this measure are
determined by $\gamma(t)<0$.

Through the above arguments, we conclude that, for an open two-level
system with damped Jaynes-Cummings model, the distributions of
non-Markovian intervals for the three measures, $\mathcal{N}$,
$\mathcal{I}^{(E)}$ and $\mathcal{I}$, are exactly the same, which
are determined uniquely by the condition $\gamma(t)<0$. According to
the viewpoint of equivalence for different measures mentioned in
Sec.II, we thus conclude that the three measures are equivalent for
detecting the existence of non-Markovianty of a two-level system
interacting with environment via Jaynes-Cummings model. In the
arguing process, we did not assume any specific structure of
spectral density. Thus the result has good adaptability. They are
valid for arbitrary structured environment (Lorentzian, Ohmic,
waveguide spectrum etc.), arbitrary detunings (including resonance)
between system and its environment, and arbitrary coupling
strengthes.

\section{The case of dephasing model}
As the second example that we use to demonstrate the equivalence of
measures of non-Markovianty, we consider a two-level atom which is
coupled to a reservoir of harmonic oscillators via dephasing model.
The Hamiltonian in Schr\"{o}dinger picture is taken to be
\begin{equation}
H=\frac{\omega_{0}}{2}\sigma_z+\sum_k{{{\omega}_k}{b_k^\dagger}{b_k}}+
\sum_k{\sigma_z({\lambda_k}{b_k^\dagger}+{\lambda_k^*}{b_k})}
\end{equation}
where $\omega_{0}$ is transition frequency and
$\sigma_{z}=|1\rangle\langle1|-|0\rangle\langle0|$ the Pauli
operator of the atom. $\omega_{k}$, $b_{k}$ are respectively the
frequency and annihilation operator for the $k$-th harmonic
oscillator of the reservoir. The coupling strength $\lambda_{k}$ is
assumed to be complex in general. This dephasing model, which is
extensively used to simulate the decoherence of a qubit coupling to
its environment in quantum information science, can be solved
exactly. For the initial state of the total system
\begin{equation}
\rho_{SB}(0)=\rho(0)\otimes\rho_B,
\end{equation}
with $\rho(0)$ the initial state of the atom and $\rho_{B}$ the
initial thermal equilibrium state of the reservoir, the evolution
for the elements of the reduced density matrix of the atom may be
written as \cite{Breuer3}
\begin{equation}
\begin{array}{l}
  \rho_{11}(t)=\rho_{11}(0),\rho_{00}(t)=\rho_{00}(0), \\
\rho_{10}(t)=\rho_{01}^*(t)=\rho_{10}(0)e^{\Gamma_{p}(t)}.
\end{array}
\end{equation}
Here the negative dephasing function $\Gamma_{p}(t)$ is defined as
\begin{equation}
\Gamma_{p}(t)=-\int_{0}^{\infty}d\omega
J(\omega)\mathrm{coth}(\omega/2k_{B}T)\frac{1-\mathrm{cos}\omega
t}{\omega^{2}},
\end{equation}
with $J(\omega)$ the spectral density of the reservoir, $k_{B}$ the
Boltzmann constant and $T$ the reservoir temperature in thermal
equilibrium. For any pair of initial states $\rho_{1}(0)$ and
$\rho_{2}(0)$ of the atom, one can easily obtain the dynamical
states $\rho_{1}(t)$ and $\rho_{2}(t)$, and then get the dynamical
trace distance as,
\begin{equation}
D(t,\rho_{1,2}(0))=\sqrt{a^{2}+|b|^{2}e^{2\Gamma_{p}(t)}}.
\end{equation}
Where
$a=\langle1|\rho_{1}(0)|1\rangle-\langle1|\rho_{2}(0)|1\rangle$ and
$b=\langle1|\rho_{1}(0)|0\rangle-\langle1|\rho_{2}(0)|0\rangle$ are
the differences of the populations and of coherences for the two
given initial states respectively. Inserting this trace distant into
eq.(1), one immediately get the non-Markovian measure for this
dephasing model as
\begin{equation}
\mathcal{N}_{p}=-2\int_{\gamma_{p(t)}<0}dt\gamma_{p}(t)\frac{|b|^{2}e^{2\Gamma_{p}(t)}}{\sqrt{a^{2}+|b|^{2}e^{2\Gamma_{p}(t)}}},
\end{equation}
where the dephasing rate is defined as
$\gamma_{p}(t)=-\frac{1}{2}\dot{\Gamma}_{p}(t)$, and based on the
same reason as before, we neglect the maximization to
$\mathcal{N}_{p}$. It shows that the non-Markovian intervals are
uniquely determined by $\gamma_{p}(t)<0$.

In order to calculate the expression of the second measure
$\mathcal{I}^{(E)}$, we introduce, in like manner as before, an
isolated ancillary atom which is initially prepared in a maximally
entangled state $|\Phi\rangle$ in eq.(12) with the system atom.
According to the evolution of eq.(26) and employing the trick in
reference \cite{Bellomo}, one can easily obtain the dynamical
density matrix for the system and ancillary atoms,
\begin{equation}
\rho_{SA}(t)=\frac{1}{2}[|10\rangle\langle10|+e^{\Gamma_{p}(t)}|01\rangle\langle10|+e^{\Gamma_{p}(t)}|10\rangle\langle01|+|01\rangle\langle01|].
\end{equation}
Where the first state refers to the system and the second one to the
ancillary atom. It is easy to find that the concurrence for this
state is $C(t)=\mathrm{exp}[\Gamma_{p}(t)]$, and thus eq.(5)
produces
\begin{equation}
\mathcal{I}^{(E)}_{p}=-4\int_{\gamma_{p}(t)<0}\gamma_{p}(t)e^{\Gamma_{p}(t)}dt.
\end{equation}
Again, the non-Markovian intervals for measure
$\mathcal{I}^{(E)}_{p}$ is also uniquely determined by
$\gamma_{p}(t)<0$.

Finally, let us calculate the measure $\mathcal{I}$ of eq.(3). It is
not difficult from eq.(26) to verify that the dynamical master
equation for the dephasing two-level atom can be written as
\begin{equation}
\frac{d\rho}{dt}=\gamma_{p}(t)[\sigma_{z}\rho\sigma_{z}-\rho],
\end{equation}
where we again have used the relation
$\dot{\Gamma}_{p}(t)=-2\gamma_{p}(t)$. Following the same deduction
for eqs.(21)-(23), we obtain
\begin{equation}
\mathcal{I}_{p}=-2\int_{\gamma_{p}(t)<0}\gamma_{p}(t)dt.
\end{equation}
Once again, the non-Markovian intervals for this measure are
uniquely determined by $\gamma_{p}(t)<0$.

All in all, for a two-level system with dephasing model, we again
draw the conclusion that the distributions of non-Markovian
intervals for the three measures are exactly the same and uniquely
determined by the condition $\gamma_{p}(t)<0$. Thus the three
measures are equivalent.
\section{Conclusion}

In conclusion, we have compared the three measures of
non-Markovianty proposed in references \cite{Breuer} and
\cite{Rivas} respectively for two-level systems interacting with
environments via damped Jaynes-Cummings or dephasing model. The
results show that the three measures have exactly the same
distribution of non-Markovian intervals, and hence are actually
equivalent each other in the aspect of detecting non-Markovianty of
quantum processes. This equivalence implies that these measures, in
different ways, well capture the intrinsical characters of
non-Markovianty of quantum processes. We have also found that the
maximization in the BLP measure can be removed for the considered
models, without influencing the sensibility of the measure to detect
non-Markovianty. This result is important which extremely avoids the
complicated mathematical calculations. The dynamical models
considered here, i.e., damped Jaynes-Cummings and dephasing models,
represent two kinds of fundamental coupling forms in studying
problems of open systems. And we did not assume any specific
spectral density for the structured environment. Thus our results
have good adaptability. They apply to arbitrary structured
environment, arbitrary detunings between system and its environment,
and arbitrary coupling strengthes. Exposing the relations between
different measures of non-Markovianty in different physical systems
and different models is important. We expect similar researches for
many more measures could be expanded for more general models or in
more complicated systems.

\section{ACKNOWLEDGMENTS}
This work is supported by the National Natural Science Foundation of
China (Grant No.11075050), the National Fundamental Research Program
of China (Grant No.2007CB925204), and the Construct Program of the
National Key Discipline.

\end{document}